# Microgrid Operation Control with State-of-Charge-Dependent Storage Power Constraints

Edgar Diego Gomez Anccas
*Chair of Electrical Power Systems*
*Helmut Schmidt University*
Hamburg, Germany
diego.gomez@hsu-hh.de

Christian A. Hans
*Automation and Sensorics*
*in Networked Systems Group*
*University of Kassel*
Kassel, Germany
hans@uni-kassel.de

Detlef Schulz
*Chair of Electrical Power Systems*
*Helmut Schmidt University*
Hamburg, Germany
detlef.schulz@hsu-hh.de

*Abstract*—The microgrid concept offers high flexibility and resilience due to the possibility of switching between grid-connected and stand-alone operation. This renders microgrids an auspicious solution for rural areas and critical infrastructure. In stand-alone or islanded mode, the main objective is cost minimization while ensuring a safe and reliable operation. Optimal operation schemes for microgrids usually assume fixed power limits for energy storage units. This, however, is not sufficient for lithium-ion energy storage systems, which often come with dynamic power limits that depend on the state of charge. These limits are especially prominent when the state of charge is close to its boundaries. In this paper, dynamic constraints for energy storages are modelled using convex polytopes and fitted to experimental data acquired from an 11.6 kWh lithium-ion energy storage system. The polytopic constraints are integrated in a model predictive control scheme that was designed for a stand-alone microgrid composed of a fuel cell, a photovoltaic generator and a lithium-ion energy storage system. To evaluate the advantages, a case study with two configurations is performed. The model predictive controller without polytopic constraints led to constraint violations in 11.77 % of the simulation time steps with a maximum deviation of 118 % above the power limits. The configuration with polytopic constraints in contrary led to no violations over the entire simulation horizon.

## I. INTRODUCTION

The transition from a few large-scale power plants to modern energy systems featuring a large number of decentralized units, of which many provide renewable energy, poses challenges at different levels [1], [2]. One solution to tackle these challenges is the segmentation of large grids into smaller microgrids (see, e.g., Fig. 1). These can be operated in parallel to the main grid or in stand-alone mode. The latter is an auspicious solution for rural areas with no connection to the main grid and for critical infrastructure to increase its resilience. To enable a reliable operation a solid control structure needs to be established which automatically and efficiently manages the units in a microgrid. Key objectives in optimal microgrid operation are maximizing economic benefits and ensuring a safe operation.

A promising approach for the operation of microgrids is model predictive control (MPC). Here, an optimization problem is solved in a receding horizon fashion, adjusting decisions as new information is gathered. Although MPC is powerful, it can be computationally expensive and is heavily reliant on suitable plant models. In microgrid operation, fixed charge and discharge power limits for energy storage units are often assumed to simplify operation control schemes [3]–[7]. This, however, can lead to issues because of dynamic voltage slopes of, e.g., lithium-ion (Li-ion) batteries, which often result in state-of-charge-dependent power limits.

Only a small number of publications have included this point in their schemes: A control strategy that adapts the primary control of an energy storage system based on its power and state-of-charge (SoC) conditions is introduced in [8]. The strategy aims to mitigate transient detrimental effects. Energy storage power limiting is implemented in [9] for a DC microgrid based on operating modes. The change of operating modes is based on renewable infeed, energy storage SoC and load conditions. However, in these two contributions no explicit optimization of the operation is implemented. In [10], SoC-dependent storage current limitation is part of an MPC scheme. In detail, the limits are part of a DC/DC converter's control cascade during grid-connected operation. However, the control scheme belongs to the converter-level control and works with prediction steps in the microsecond range with different objectives than operation control. Time of use tariffs may lead to simultaneous charging of multiple storage devices and can thus cause a violation of distribution network constraints. To better allocate network capacity an independent storage operator is introduced in [11]. The storage operator manages power limits of active energy storages based on predefined affine functions aiming to keep the distribution network at safe operation and increase customer profits. In [12], dynamic power constraints are introduced that are defined in circular segments which vary according to the SoC as part of a rule-based energy management system aiming to improve the systems' transient performance under various operation conditions. In summary, although the present literature gives valuable insights, it focuses on different optimization schemes, time regions, control layers and topologies. Therefore, existing approaches cannot be used for microgrid operation control.

This paper introduces a method to describe the SoC-dependent power limits of energy storage units using convex polytopes. These can be formed by sets of affine constraints and are therefore easily usable in MPC schemes. The resulting constraints prevent infeasible power setpoints, which may jeopardize a safe operation. Overall, the contributions of this work are as follows:

1) Derivation of dynamic, SoC-dependent power constraints for energy storage units based on convex polytopes and definition of these constraints as a set of affine inequalities.
2) Forming such a convex polytope from experimental data from an 11.6 kWh Li-ion storage unit which is connected to the grid through a 15 kW inverter.
3) Integration of the fitted dynamic constraints in a prescient MPC scheme for a stand-alone microgrid model.
4) Assessment of the advantages of the novel operation control scheme through numerical simulations with and without the polytopic constraints.

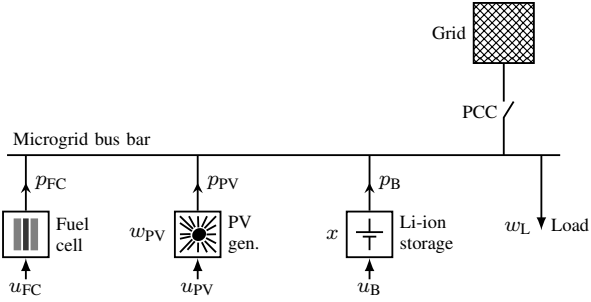

Fig. 1. Microgrid under investigation. Image motivated by [13]

The remainder of this paper is structured as follows. In Section II, the model of a microgrid in stand-alone operation including dynamic SoC-dependent power constraints is presented. Then, in Section III, the cost function is derived and a prescient MPC scheme is introduced. In Section IV, the scheme is employed in closed-loop simulations for two different configurations. The results are then summarized in Section V.

*A. Notation*

The sets of nonnegative integers, real numbers, positive real numbers, nonnegative real numbers and negative real numbers are $\mathbb{N}$, $\mathbb{R}$, $\mathbb{R}_{>0}$, $\mathbb{R}_{\geq 0}$ and $\mathbb{R}_{<0}$, respectively. The set of Booleans is $\mathbb{B} = \{0, 1\}$. A forecast of variable $a$ at time $k \in \mathbb{N}$ for a future timestep $k + j \in \mathbb{N}$ is described by $a(k+j|k)$. The vector with elements $a_i, \ldots, a_N$ is described by $[a_i]_{i=1}^{N} = [a_i \cdots a_N]^\top$ with $N \in \mathbb{N}$.

## II. MICROGRID MODEL

The system under investigation is a low voltage AC microgrid consisting of four units which can be connected to the grid through a point of common coupling (PCC) as depicted in Fig. 1. While a connection to the main grid is possible, the investigations in this work are focused on islanded operation of the microgrid. The fuel cell (FC) system and the battery storage are both operated in grid-forming mode and droop controlled [14]. The third unit is a maximum power point tracking-controlled photovoltaic (PV) generator. A three phase AC load is also an element of the microgrid. The system is modeled along the lines of [15] and subsequently extended to include dynamic storage constraints. We consider a discrete time model with sampling time $\Delta T \in \mathbb{R}_{>0}$ and time instants $k \in \mathbb{N}$.

*A. PV system*

The PV system is bounded by the device's maximum $p_{\text{PVmax}} \in \mathbb{R}_{>0}$ and minimum output power $p_{\text{PVmin}} \in \mathbb{R}_{\geq 0}$, i.e.,

$$p_{\text{PVmin}} \leq p_{\text{PV}}(k) \leq p_{\text{PVmax}}, \quad (1a)$$
$$p_{\text{PVmin}} \leq u_{\text{PV}}(k) \leq p_{\text{PVmax}}. \quad (1b)$$

A curtailment of the infeed of renewable energy is taken into account, by using the PV power setpoint $u_{\text{PV}}(k) \in \mathbb{R}_{\geq 0}$. It can be chosen to reduce the infeed below the available PV power given by the uncertain input $w_{\text{PV}}(k) \in \mathbb{R}_{\geq 0}$. Thus, the curtailed PV output power is

$$p_{\text{PV}}(k) = \min\left(u_{\text{PV}}(k), w_{\text{PV}}(k)\right). \quad (2)$$

Using big-M reformulations [16], this constraint can be transformed into a set of affine inequalities, i.e.,

$$u_{\text{PV}}(k) - M_{\text{PV}} \cdot \delta_{\text{PV}}(k) \leq p_{\text{PV}}(k) \leq u_{\text{PV}}(k), \quad (3a)$$
$$w_{\text{PV}}(k) - m_{\text{PV}} \cdot (1 - \delta_{\text{PV}}(k)) \leq p_{\text{PV}}(k) \leq w_{\text{PV}}(k), \quad (3b)$$

with Boolean decision variable $\delta_{\text{PV}}(k) \in \mathbb{B}$ and big-M parameters $M_{\text{PV}} \in \mathbb{R}_{>0}, m_{\text{PV}} \in \mathbb{R}_{<0}$, where $m_{\text{PV}} < p_{\text{PVmin}} - p_{\text{PVmax}}$ and $M_{\text{PV}} > p_{\text{PVmax}} - p_{\text{PVmin}}$.

*B. Fuel cell*

Power $p_{\text{FC}} \in \mathbb{R}_{\geq 0}$ and setpoints $u_{\text{FC}} \in \mathbb{R}_{\geq 0}$ are limited by the constraints

$$\delta_{\text{FC}}(k) \cdot p_{\text{FCmin}} \leq p_{\text{FC}}(k) \leq p_{\text{FCmax}} \cdot \delta_{\text{FC}}(k), \quad (4a)$$
$$\delta_{\text{FC}}(k) \cdot p_{\text{FCmin}} \leq u_{\text{FC}}(k) \leq p_{\text{FCmax}} \cdot \delta_{\text{FC}}(k), \quad (4b)$$

with $p_{\text{FCmin}} \in \mathbb{R}_{\geq 0}$ and $p_{\text{FCmax}} \in \mathbb{R}_{>0}$. Whether the FC is on or off is indicated by the Boolean decision variable $\delta_{\text{FC}}(k) \in \mathbb{B}$. Note that (4) implements the condition that if $\delta_{\text{FC}}(k) = 0$, then $p_{\text{FC}}(k) = u_{\text{FC}}(k) \stackrel{!}{=} 0$; and if $\delta_{\text{FC}}(k) = 1$, then $p_{\text{FC}}(k) \in [p_{\text{FCmin}}, p_{\text{FCmax}}]$ and $u_{\text{FC}}(k) \in [p_{\text{FCmin}}, p_{\text{FCmax}}]$.

*C. Battery storage system*

Li-ion batteries show SoC-dependent voltage characteristics that can be segmented into three zones as illustrated in Fig. 2. At low SoC levels (Region I), the Li-ion concentration at the anode depletes, decreasing the potential difference to the lithiated cathode. Additionally, the internal resistance increases at low SoC levels due to slower reaction kinetics. Therefore, the discharge power is often limited in region 1 to avoid power levels detrimental to battery longevity. This includes a limitation of charging currents to avoid additional heat generation in this high resistance regime. At intermediate SoC levels (Region II in Fig. 2), the Li-ion concentration and thus the battery voltage remains in a narrow range. In this region the battery has no further restrictions beyond its maximum charge and discharge power. At high SoC levels (Region III in Fig. 2), the potential difference between the depleted cathode and the saturated anode increases. This region is also characterized by an increased internal resistance. The charging current in this region is limited due to heat generation and proximity to the maximum battery voltage. To take these effects into consideration, manufacturers often include SoC-dependent power restrictions such as those shown in Fig. 3. The black graphs illustrate the power limits acquired from a 422 V, 11.6 kWh Li-ion storage during operation. The vertical lines, $x_{\min} \in \mathbb{R}_{>0}$ and $x_{\max} \in \mathbb{R}_{>0}$, indicate the system's SoC boundaries. The horizontal line $p_{B\min} \in \mathbb{R}_{<0}$ sets the charging limits of the battery, while the limit $p_{B\max} \in \mathbb{R}_{>0}$ represents the upper discharging restriction imposed by the battery inverter, which is smaller than the battery power limits at some regions. Power and energy values within the nonconvex area enclosed by the black graphs as well as the vertical and horizontal lines are allowed. For use in convex MPC, the admissible battery power $p_B$ and energy $x$ can be approximated by a convex polytope of the form

$$\mathbb{P} = \left\{ (x, p_B) \,\middle|\, \begin{array}{l} x \in [x_{\min}, x_{\max}] \subset \mathbb{R}, \\ p_B \in [p_{B\min}, p_{B\max}] \subset \mathbb{R}, \\ p_B \leq \alpha_i x + \beta_i, \forall i \in \mathbb{L} \end{array} \right\}. \quad (5)$$

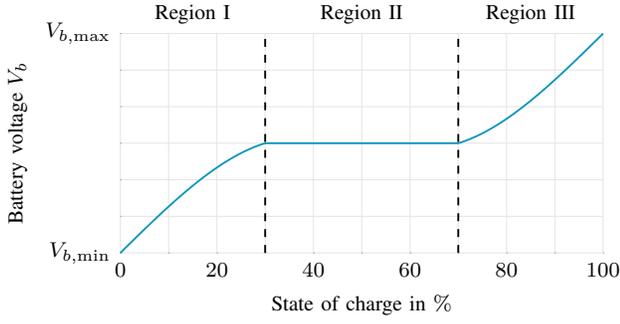

Fig. 2. SoC-dependent open-circuit voltage of a Li-ion battery.

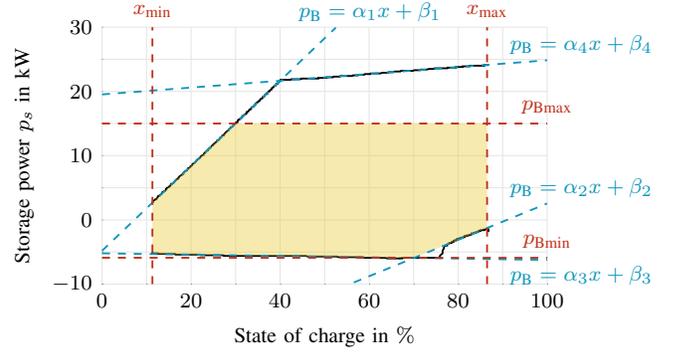

Fig. 3. Feasible area for storage power. Fitting the polytope to experimental data measured in a 422 V, 11.6 kWh Li-ion battery system yields the parameters $\{\alpha_1, \ldots, \alpha_4\} = \{0.66, 0.28, -0.01, 0.05\}$, as well as $\{\beta_1, \ldots, \beta_4\} = \{-4.81, -25.77, -5.21, 19.53\}$ and the intervals $[x_{\min}, x_{\max}] = [11.31, 86.5]$ as well as $[p_{B\min}, p_{B\max}] = [-5.9, 15]$.

The inequalities at the bottom describe SoC-dependent power limit which are formed using the parameters $\alpha_i \in \mathbb{R}$, $\beta_i \in \mathbb{R}$ for $i \in \mathbb{L} \subset \mathbb{N}$.

*Remark* 1. The measurement data was fitted to a polytope by finding the optimal intersection points that yielded the minimum least squares error. It is important to note that different energy storages might need additional inequalities to model.

Using (5) allows us to model the battery constraints as

$$(x(k), p_B(k)) \in \mathbb{P}, \quad (6a)$$
$$(x(k+1), p_B(k)) \in \mathbb{P}, \quad (6b)$$
$$(x(k), u_B(k)) \in \mathbb{P}, \quad (6c)$$
$$(x(k+1), u_B(k)) \in \mathbb{P}. \quad (6d)$$

Note that in (6b) and (6d), the future state $k+1$ is included to make sure that limitations from all state values in the interval $[k, k+1]$ are accounted for.

Finally, the battery system's SoC dynamics are

$$x(k+1) = x(k) - \Delta T \cdot p_B(k). \quad (7)$$

### D. Power sharing

Differences between power setpoints and loads can cause frequency deviations. For grid-forming unit $i$ in a microgrid, the droop control law according to [17] reads

$$\dot{\delta}_i = \omega_i, \quad (8a)$$
$$\tau_i \dot{\omega}_i = k_i(p_i - u_i). \quad (8b)$$

Here $\dot{\delta}_i$ is the phase angle and $\omega_i$ the frequency of unit $i$. Moreover, $k_i \in \mathbb{R}_{>0}$ is the droop gain and $\tau_i$ the low pass filter time constant. At steady state, the system exhibits synchronized motion and $\omega_i = \omega_j$ [18]. Therefore, proportional power sharing for enabled units of the form

$$k_i(p_i - u_i) = k_j(p_j - u_j) \quad (9)$$

holds. For our example microgrid in Fig. 1 and using an auxiliary variable $\mu(k)$, we can express (9) as

$$k_B(p_B - u_B) = \mu(k), \quad (10)$$
$$k_{FC}(p_{FC} - u_{FC}) = \mu(k). \quad (11)$$

If a unit is not enabled, then it cannot participate in power sharing. This is accounted for by the constraint

$$k_{FC}(p_{FC}(k) - u_{FC}(k)) = \mu(k)\delta_{FC}(k). \quad (12)$$

Using the big-M reformulation, this can be transformed into

$$k_{FC}(p_{FC} - u_{FC}) \leq M_{FC} \cdot \delta_{FC}(k), \quad (13a)$$
$$k_{FC}(p_{FC} - u_{FC}) \geq m_{FC} \cdot \delta_{FC}(k), \quad (13b)$$
$$k_{FC}(p_{FC} - u_{FC}) \leq \mu(k) - m_{FC} \cdot (1 - \delta_{FC}(k)), \quad (13c)$$
$$k_{FC}(p_{FC} - u_{FC}) \geq \mu(k) - M_{FC} \cdot (1 - \delta_{FC}(k)). \quad (13d)$$

## E. Power equilibrium

The power equilibrium constraint

$$p_{FC}(k) + p_B(k) + p_{PV}(k) = w_L(k), \quad (14)$$

with $w_L(k)$ as the load completes the microgrid model.

## III. PROBLEM FORMULATION

### A. Cost function

The overall cost is divided into different parts to consider the system and its devices comprehensively. The cost of renewable operation $l_r$ rewards renewable infeed using

$$l_r(k) = c''_{PV} \cdot (p_{PVmax} - p_{PV}(k))^2 \quad (15)$$

The fuel cell costs are separated into running osts $l_{FCru}$ and switching costs $l_{FCsw}$ i.e.,

$$l_{FCru}(k) = c_{FCru} \cdot \delta_{FC}(k) + c'_{FCru} \cdot p_{FC}(k), \quad (16)$$
$$l_{FCsw}(k) = c_{FCsw} \cdot (\delta_{FC}(k) - \delta_{FC}(k-1))^2. \quad (17)$$

The battery costs are composed of costs related to conversion losses i.e.,

$$l_B = c''_B \cdot p_B(k)^2. \quad (18)$$

Finally, we can calculate the overall cost as

$$l(k) = l_{FCru}(k) + l_{FCsw}(k) + l_r(k) + l_B(k). \quad (19)$$

### B. Prescient MPC

With the three vectors

$$v(k) = [u_{FC}(k) \quad u_B(k) \quad u_{PV}(k) \quad \delta_{FC}(k)],$$
$$z(k) = [p_{FC}(k) \quad p_B(k) \quad p_{PV}(k) \quad \delta_{FC}(k) \quad \mu(k)],$$
$$w(k) = [w_{PV}(k) \quad w_L(k)],$$

and decision variables

$$\mathbf{v} = v[(k+j|k)]_{j=0}^{J-1},$$
$$\mathbf{x} = x[(k+j|k)]_{j=0}^{J},$$
$$\mathbf{z} = z[(k+j|k)]_{j=0}^{J-1},$$

the prescient MPC with discount factor $\gamma \in (0,1) \subset \mathbb{R}$ reads as follows.

*Problem* 1 (Prescient MPC). Solve

$$\min_{\mathbf{v},\mathbf{x},\mathbf{z}} \sum_{j=0}^{J-1} l(k+j|k) \cdot \gamma^{j+1}$$

subject to

(1), (3)–(7), (10), (13)–(14), $\forall j = 0, \ldots, J-1$
with initial conditions, $x(k|k) = x(k), \delta_{FC}(k|k) = v(k-1)$.

As shown in Fig. 4, forecasts up to the prediction horizon $J \in \mathbb{N}$ as well as the measurements for the initial conditions are fed into the MPC, where Problem 1 is solved. The resulting optimal setpoints $v^*(k)$ are then applied to the microgrid, which runs and after some time provides the next state determined from the setpoints and $w(k)$. The new state $x(k)$ and previous input $v(k-1)$ are sent to the controller and the problem is solved repeatedly in a receding horizon fashion.

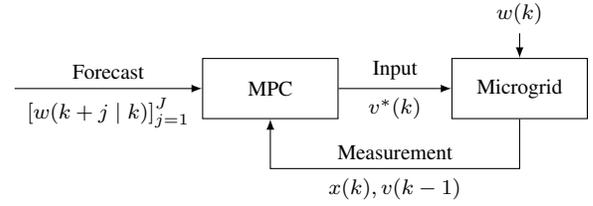

Fig. 4. MPC scheme for time instant k

## IV. CASE STUDY

As the focus of this work lies in modelling differences, a prescient MPC is employed to neglect prediction errors. This implies that the forecasts $w[(k+j|k)]_{j=0}^{J}$ and uncertain input $w(k)$ are identical at each timestep. SoC-dependent constraints in an MPC scheme can improve performance and help to avoid infeasible setpoints. To assess this claim, the microgrid and MPC scheme introduced in Section III are employed in two configurations. In the first configuration, the polytopic constraints (6) are not included in the MPC scheme, whereas in the second one, they are part of the MPC scheme.

In order to map realistic responses, in both scenarios the limitation (6a) is part of the low-level microgrid controls in time instants where the FC is enabled. If the FC is disabled, then the battery has to provide all power. However, this over/under power operating condition is potentially dangerous and will therefore be marked and analysed in the following case study. Furthermore, to account for dynamic SoC-dependent power limits the MPC and microgrid model are simulated with different sampling times (see Table I). For the case study, load data from [19] was used and scaled appropriately. The irradiance data $\hat{I}(k)$ is acquired from [20] and the available PV infeed is derived from

$$w_{PV}(k) = \begin{cases} 0, & \text{if } \hat{I}(k) < 0\,\text{W/m}^2, \\ p_{PVmax}, & \text{if } \hat{I}(k) > 1000\,\text{W/m}^2, \\ \frac{\hat{I}(k)}{1000} p_{PVmax}, & \text{else.} \end{cases} \quad (20)$$

The simulation parameters for both configurations are portrayed in Table I.

*Remark* 2. A parameter sensitivity analysis showed that $p_{PVmax} > w_L > p_{Bmax}$ leads to constraint violations. Therefore,

TABLE I
SIMULATION PARAMETER

| Variable | Interval/Value | Variable | Value |
|---|---|---|---|
| $p_B$ | [-0.59, 1.5] | Prediction horizon $J$ | 6 |
| $p_{FC}$ | [0.2, 4.5] | Simulation duration | 48:00 h |
| $p_{PV}$ | [0, 4.5] | Sample time MPC | 00:30 h |
| | | Sample time MG | 00:01 h |
| $x$ | [0.258, 1.972] h | Base pu | 10 kW |
| $c''_{PV}$ | 1 | $c_{FCru}$ | 0.13 |
| $c'_{FCru}$ | 4.56 | $c_{FCsw}$ | 0.2 |
| $c''_B$ | 0.1 | $\gamma$ | 0.9 |
| $\delta_{FC}(0)$ | 0 | $x(0)$ | 1.5 |

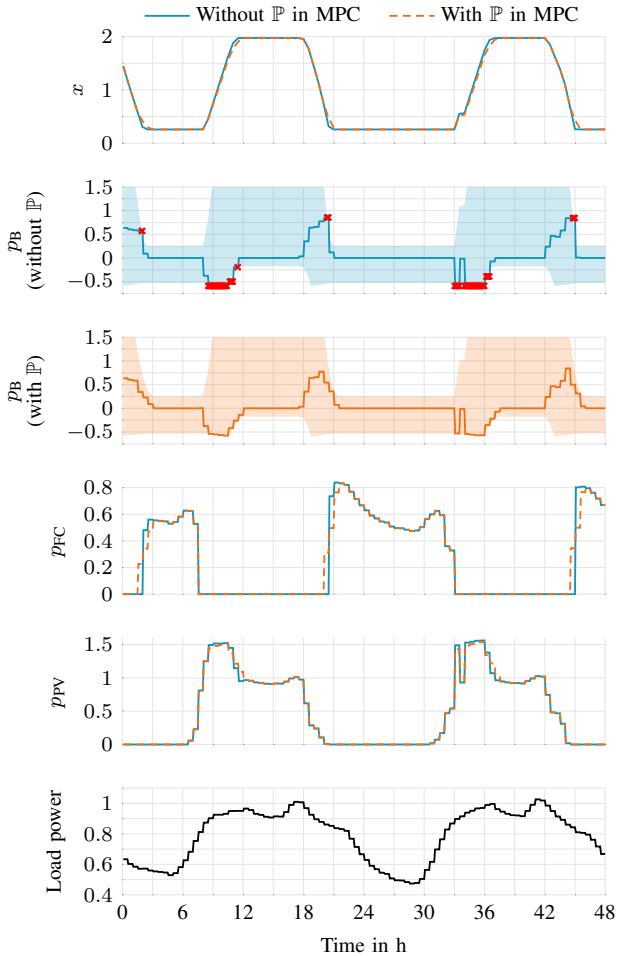

Fig. 5. Power and energy trajectories of the elements in the stand-alone microgrid in pu for the two configurations. The red crosses indicate instances, where constraint violations occur.

the listed intervals in Table I are chosen for the case study. Note that this represents a very relevant scenario in future power systems.

The simulations are performed over duration of 48 h, yielding the data shown in Fig. 5. Here, the energy $x$ of both configurations differs in regions close to their boundaries $x_{\min}$ and $x_{\max}$. At these points, the battery power exceeds the power limits, represented by the shaded area, for the configuration without $\mathbb{P}$. The constraint violations are highlighted by red crosses at their points of occurrence. For the MPC with $\mathbb{P}$, no violation of the power limits occur. The constraints smooth the trajectory of $x$ by triggering the activation of the FC ($\delta_{\mathrm{FC}} = 1$) early, thereby avoiding power limit violations during discharging near the lower energy bounds. In contrast, neglecting $\mathbb{P}$ in the MPC results in an overestimation of the available $p_{\mathrm{B}}$ leading to repeated constraint violations at these points. Furthermore, in regions where $p_{\mathrm{PV}} > w_{\mathrm{L}}$, the FC is deactivated ($\delta_{\mathrm{FC}} = 0$) and the battery switches to charging. As the charging power limits are tighter, the $p_{\mathrm{B}}$ setpoints lead to repeated constraint violations at the upper and lower energy bounds for the MPC without $\mathbb{P}$.

In real-world applications current limitations in the inverter would shield the battery from power setpoints beyond its limits thus causing power imbalances. These imbalances depend on the magnitude of the limit breach. As only the configuration without $\mathbb{P}$ exhibits breaches, their absolute magnitude and count is illustrated in Fig. 6. The amount of time that violations occur adds up to 339 minutes, which corresponds to more than 10 % of the time. The deviations from the threshold vary in magnitude with more than 85 % being below 0.1 p.u. However, some deviations are higher and peak at 0.58 p.u. Such a deviation would cause a system shutdown (for the sake of demonstration, the simulation is continued). Although most of the constraint violations are not of a magnitude causing a system shutdown, they would still leave the system in an undesired state where the system is vulnerable to further disturbances. For the harshest constraint violation, the deviations would lead to a system shutdown.

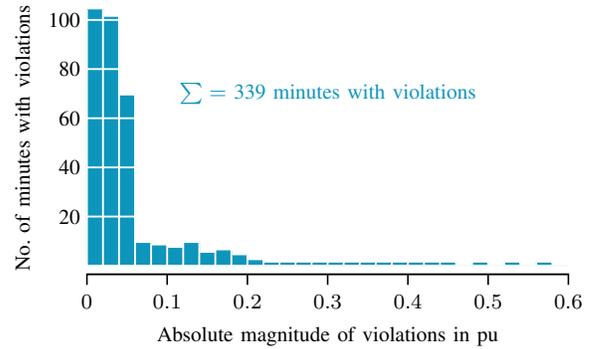

Fig. 6. Violations during simulations

## V. CONCLUSIONS

In microgrid operation control, optimization schemes customarily assume energy storage power limits to be constant. This simplifying assumption does not hold for Li-ion energy storages, which exhibit increasingly restrictive power limits as they approach their SoC bounds. Therefore, this work introduces a method to approximate said dynamic constraints by convex polytopes. The polytopic constraints can be easily integrated in MPC schemes through a set of affine inequalities. To evaluate performance improvements due to the constraints, 48 h simulations were performed for two MPC schemes. For the configuration using the constraints, security is dramatically improved as no constraint violations occur. In contrast, the configuration neglecting the constraints led to multiple violations in more than 10 % of the simulation time. The harshest constraint violation with a maximum deviation of 0.58 p.u, would endanger system operation in real-world applications. These results emphasize the benefits of including the dynamic power constraints.

Future work will involve applying the proposed method to the laboratory environment presented in [21] and [22].

ACKNOWLEDGMENT

This research paper is funded by dtec.bw–Digitalization and Technology Research Center of the Bundeswehr–which we gratefully acknowledge. dtec.bw is funded by the European Union–NextGenerationEU.